\pdfoutput=1
\documentclass[conference]{IEEEtran}
\usepackage[utf8]{inputenc}
\usepackage{url}
\usepackage{graphicx}

\begin{document}

\title{Closing the Blinds: Four Strategies for Protecting Smart Home Privacy from Network Observers}

\author{\IEEEauthorblockN{Noah Apthorpe, Dillon Reisman,  Nick Feamster}

\IEEEauthorblockA{Computer Science Department\\
Princeton University\\
apthorpe@cs.princeton.edu, dreisman@princeton.edu, feamster@cs.princeton.edu}}

\maketitle

\begin{abstract}
The growing market for smart home IoT devices promises new conveniences for consumers while presenting novel challenges for preserving privacy within the home.
Specifically, Internet service providers or neighborhood WiFi eavesdroppers can measure Internet traffic rates from smart home devices and infer consumers' private in-home behaviors.
Here we propose four strategies that device manufacturers and third parties can take to protect consumers from side-channel traffic rate privacy threats: 
1) blocking traffic, 2) concealing DNS, 3) tunneling traffic, and 4) shaping and injecting traffic. 
We hope that these strategies, and the implementation nuances we discuss, will provide a foundation for the future development of privacy-sensitive smart homes.
\end{abstract}

\section{Introduction}
\label{sec:intro}
Internet-connected physical devices have rapidly increased in popularity and commercial availability within the past several years.  
This trend, called the Internet of Things (IoT), includes many consumer products sold to replace traditional non-networked home appliances or introduce novel technologies into consumer homes.  
These devices promise many advantages, including reduced energy consumption,  more effective health management, and living spaces that react adaptively to fit consumers' lifestyles. 

However, a future with ``smart'' homes containing numerous Internet-connected devices raises substantial privacy concerns.
In earlier work, we demonstrated that a passive network observer can infer sensitive information about consumers from the network behavior of their smart home devices, even when those devices use encryption \cite{datpaper2016}. 
Such observers include Internet service providers (ISPs), WiFi eavesdroppers, or state-level surveillance entities.
Even without the contents of network traffic, observers can use Domain Name System (DNS) queries to identify smart home devices with potential ramifications for consumer privacy.
For example, learning that a consumer has an IoT blood sugar monitor or pacemaker effectively reveals a diabetes or heart-disease diagnosis, respectively.  

Once observers identify particular smart home devices, they can then use traffic rates from identified devices to infer potentially private consumer activities.
By testing several commercially-available devices, we found that changes in traffic rates closely correlated to consumer behaviors.
For example, specific changes in traffic from a sleep monitor revealed consumer sleep patterns \cite{datpaper2016}.

Policymakers are already concerned with known security and privacy issues in smart home devices and have solicited ideas for how to protect consumers \cite{ftcchallenge}. 
Our results present a new class of side-channel privacy threats that policymakers and engineers will have to take into account. 
In this paper, we present strategies for addressing these threats and evaluate their effectiveness.
Each proposed strategy seeks to prevent device identification or behavior inference by limiting an observer's access to or confidence in data collected from a smart home.
The strategies could be deployed on individual devices or on third-party hubs or routers.

We consider four strategies in light of two types of observers: a last-mile network observer (such as an ISP) and a WiFi eavesdropper:

\begin{enumerate}
\item Blocking outgoing connections to deprive an observer of smart home device data.

\item Encrypting DNS queries to prevent an observer from identifying devices.

\item Tunneling all smart home traffic through a virtual private network (VPN), preventing an observer from correlating tunneled traffic  originating from a smart home to individual devices. 

\item Shaping or injecting traffic to limit an observer's confidence when identifying devices or inferring behaviors, either by masking interesting traffic patterns or spoofing devices that are not on the network.

\end{enumerate}

Consumer privacy in the context of IoT smart homes has sparked considerable interest in industry, government, and the general public. 
Given that IoT traffic rate metadata is an understudied threat, we hope that the discussion and recommendations in this paper  inform the design and deployment of IoT privacy solutions.

\section{Adversarial Models}
\label{sec:threat}
Our analysis involves two adversarial models relevant to the privacy concerns of smart home device owners.

The first model is a last-mile passive network observer.
This observer has the ability to collect and analyze traffic from the external interface of a smart home gateway router.
It has no access to local area network (LAN) traffic from within the home network nor can it manipulate traffic.
The ISP that operates the last-mile connection to a consumer's smart home is a concrete example of this type of adversary. 
 
The second model is a WiFi eavesdropper.
An adversary in this model is located close enough to a smart home's WiFi network to intercept WiFi radio transmissions, but does not have the credentials to associate with the smart home's WiFi access point.
This prevents the adversary from reading encrypted packet contents after 802.11 headers (which include MAC addresses).
A neighbor in an adjacent apartment or a discrete radio receiver placed near a smart home are real-world examples of this type of adversary. 

Despite their differences,  these two adversaries both have access to traffic rate metadata. 
Each adversary can also associate traffic with individual devices  using MAC addresses (for the WiFi eavesdropper) or IP addresses (for the last-mile observer). This is possible even if the adversary has not identified device types.  

We assume that neither adversary has access to packet contents. 
In the case of the last-mile network observer, many smart home IoT devices follow best practices and use SSL encryption to protect application-layer contents \cite{datpaper2016}.
The WiFi eavesdropper cannot read IP or transport layer headers encrypted by the WPA2 protocol on password-protected access points. 

\section{Traffic Analysis Threat}

Here we provide an overview of the techniques employed in our previous research to learn sensitive consumer information from metadata of encrypted smart home device traffic \cite{datpaper2016}.

\subsection{Device identification}
An adversary observing network traffic from a consumer smart home can separate the traffic into streams and associate each stream with an individual device. 
The simplest way to do this is to group traffic by the IP addresses of the external services with which the devices are communicating.
DNS queries for domain names corresponding to these IP addresses often contain the name of the device manufacturer or even the make of the device itself. 
A last-mile observer can use these unencrypted DNS queries to identify the type of devices generating the traffic.

This approach is not possible for a WiFi eavesdropper, because DNS queries are encrypted in WiFi radio frames. 
Although DNS queries make identification easy, we demonstrate in this paper that DNS queries are ultimately unnecessary. 
Instead, an adversary can identify devices using simple supervised machine learning on traffic rate metadata (\S~\ref{sec:DNSconceal}).

\subsection{Behavior inference}
Learning the identity of individual smart home devices allows the adversary to infer sensitive consumer behaviors from changes in traffic patterns.
Given the limited-purpose nature of many IoT devices (e.g., a smart power outlet's only function is turning on or off), it is often straightforward for an adversary to map changes in traffic rates to a particular consumer actions.
We examined a commercially-available IoT sleep monitor, personal assistant, security camera, and smart power outlet and found that each allowed an adversary to easily infer consumer behaviors from traffic patterns.
We believe that this type of side-channel privacy attack using traffic rate metadata is applicable to a broad range of IoT smart home devices. 

\section{Strategies for Smart Home Privacy Preservation}
\label{sec:solutions}
Here we present four technical strategies to protect consumers from smart home traffic rate metadata privacy threats. 
We evaluate the effectiveness of these strategies under last-mile observer and WiFi eavesdropper adversary models and compare their feasibility with or without device manufacturer support. We consider smart homes with a variety of WiFi  devices from multiple manufacturers.

\subsection{Blocking traffic}
One strategy for preventing device identification and consumer behavior inference is to prevent an adversary from collecting smart home network traffic in the first place. 
The most naive method is to deploy a firewall that prevents all IoT device traffic from leaving a smart home's local network. 
Configuring such a firewall is straightforward on an intermediate hub-like device or the home gateway router.
However, the success of this approach depends on the extent to which devices retain their functionality when blocked from the Internet.

We tested the effect of removing Internet connectivity from seven commercially available devices  while maintaining the local network (Figure~\ref{fig:blocking}). 
Four of the devices retained limited functionality, although many ``smart" features which distinguish these devices from their non-IoT analogs were lost.
The remaining three devices had no functionality without Internet connection and were completely unusable.
This was somewhat unexpected, because there is no technical reason why these devices could not have continued to partially function.
The sleep monitor still could have reported current sleeping conditions detected using on-device sensors. 
The security cameras could have allowed a smartphone also on the local network to view the video feed (e.g., for monitoring an infant while in another room). 

\begin{figure*}
\begin{center}
\small
\begin{tabular}{l|l|l}
\textbf{Device} & \textbf{Functionality} & \textbf{Description} \\
\hline
Amazon Echo & limited & Can use as a bluetooth speaker with previously paired smartphone \\
&& Echo recognizes ``Alexa" keyword but does not provide any voice-control features \\ \hline
Belkin WeMo Switch & limited & Can turn switch on/off with physical button on device \\ 
&& Cannot use smartphone app to control device even when phone on local network\\ \hline
Orvibo Smart Socket & limited & Can turn switch on/off with physical button on device or smartphone app on local network \\ \hline
TP-Link Smart Plug & limited & Can turn switch on/off with physical button on device or smartphone app on local network \\ \hline
Nest Security Camera & none & Unable to view video feed or receive detected motion notifications \\ \hline
Amcrest Security Camera & none & Unable to view video feed or control camera direction\\ \hline
Sense Sleep Monitor & none & Monitor does not record sleep data\\&& Light-based UI does not reflect local sensor readings \\
&& Cannot use smartphone app to control device or access current data 
\end{tabular}
\caption{All tested commercially-available IoT devices had limited or no functionality when firewalled to prevent communication outside of the smart home LAN. This suggests that device manufacturers should be encouraged to improve their ``minimum reliable product.'' }
\label{fig:blocking}
\end{center}
\end{figure*}

These results suggest that many device manufacturers fail to include in the design of their device a strong ``minimum reliable product'' \cite{minimumreliableproduct}.
A smart device's ``minimum reliable product'' is its basic functionality in the absence of cloud support.
For our purposes, a strong minimum reliable product could allow a user to enjoy their smart device while blocking privacy-compromising network communications.
In general, providing minimal device functionality in the absence of the Internet makes products more robust.
For example, a smart thermostat should, at the very least, match the functionality of a traditional thermostat if its server goes down.
In one case, Nest users lost heat in their homes when their thermostats failed to connect to Nest servers---an obviously poor user experience \cite{nestbroken}.
For these reasons, policymakers should encourage device manufacturers to consider their device's minimum reliable product during development.

Rather than blocking all traffic indiscriminately, a hub or router could use more advanced analysis to discern what specific traffic from each device can be blocked safely.
If done correctly, smart home devices could function effectively while particularly sensitive traffic streams are prevented from leaving the home.
One possible approach is for device developers to intentionally separate information sent to cloud servers into multiple streams categorized by information type and/or necessity for device functionality. 
A firewall could then block specific streams to trade-off consumers' privacy versus usability preferences. 

There may be cases where a developer is unable to separate streams or is disincentivized to allow the consumer to block specific streams.
Without developer support, a third-party hub or router could potentially learn what traffic from particular devices can be blocked without excessively limiting functionality. 
However, encrypted traffic contents and the potential for delayed device failures makes this an open research question. 

Blocking traffic  will not prevent a WiFi eavesdropper from collecting WiFi frames between IoT devices and the home gateway router. 
Nevertheless, the development of functionality-preserving methods for blocking specific traffic streams would be a major step forward for protecting private consumer behavior from last-mile observers. 

\subsection{Concealing DNS}
\label{sec:DNSconceal}
DNS queries from smart home devices often contain or provide clues to the name of the device manufacturer or the device type itself \cite{datpaper2016}. 
Last-mile observers can therefore use DNS to identify devices inside a smart home. 
Masking DNS queries could prevent device identification by last mile observers, limiting their ability to infer consumer behaviors from traffic rates.
This would be similar to the limits that already exist on Wifi eavesdroppers.
WiFi eavesdroppers do not have access to DNS queries because they are contained in an inner layer of an encrypted 802.11 frame.

\begin{figure}[t]
\centering
\includegraphics [width=0.48\textwidth]{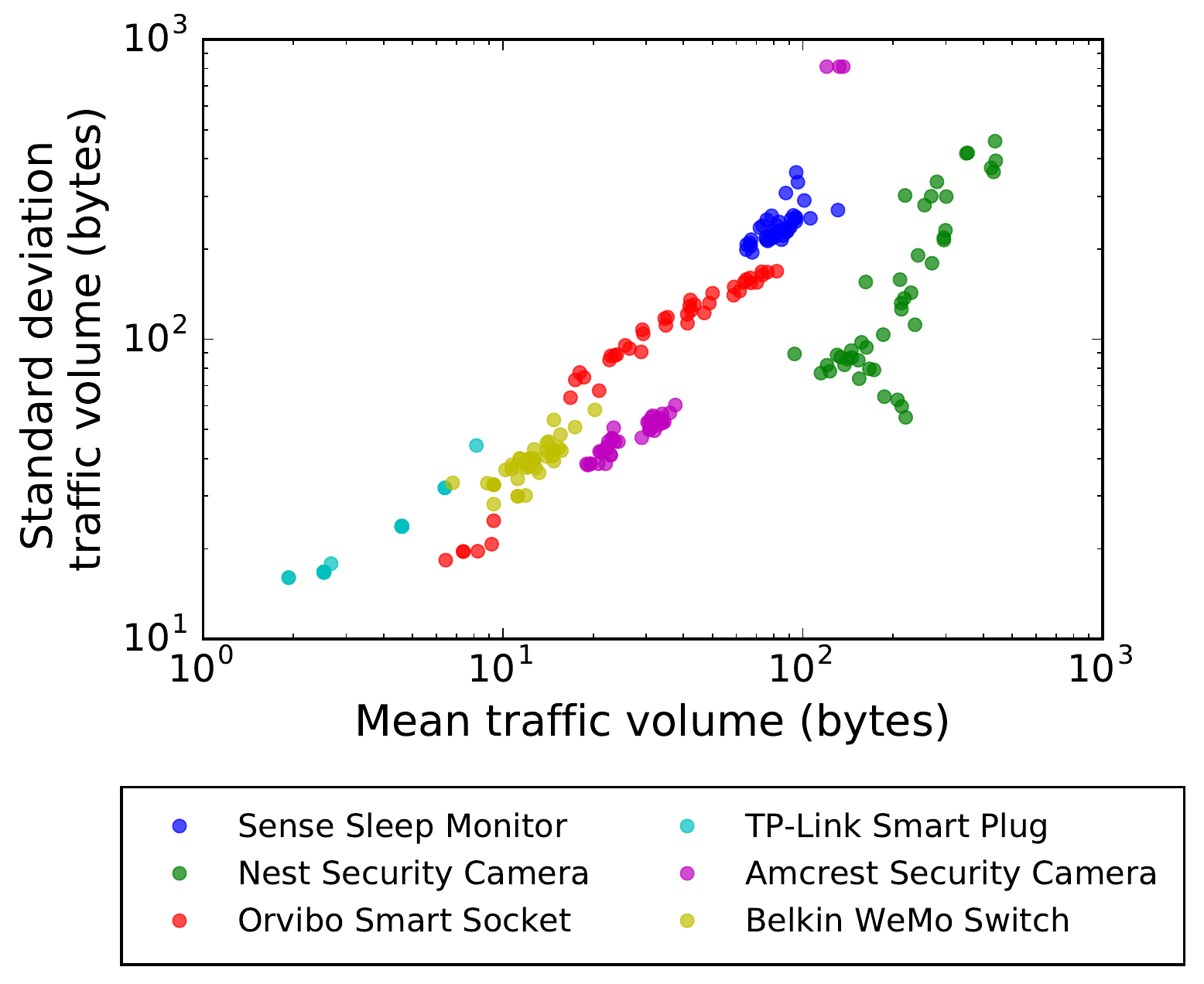}
\caption{Traffic to and from individual smart home devices clusters well by the mean and standard deviation of traffic volume within a several-minute time window. This indicates that device identification is possible using machine learning on traffic rates without additional categorical information (e.g., DNS queries).}
\label{fig:clustering}
\end{figure}

However, we have found that while preventing access to DNS queries makes device identification more difficult, it is still possible to identify devices using simple supervised machine learning on device traffic rates. 
We collected two hours worth of network traffic from six smart home IoT devices operating with minimal user interactions.  
This traffic was separated into sets corresponding to individual devices and converted into time series vectors. 
Vector elements contained the amount of traffic (bytes) sent or received by the device in consecutive \mbox{$s$-second} samples. 
These vectors were further divided into \mbox{$w$-second} windows. 
Windows were processed into \mbox{2-element} feature vectors containing the mean and standard deviation of the traffic amounts in the elements of each window.  
These simple feature vectors clustered well by the device responsible for the traffic (Figure~\ref{fig:clustering}).

\begin{figure}[t]
\centering
\includegraphics[width=0.48\textwidth]{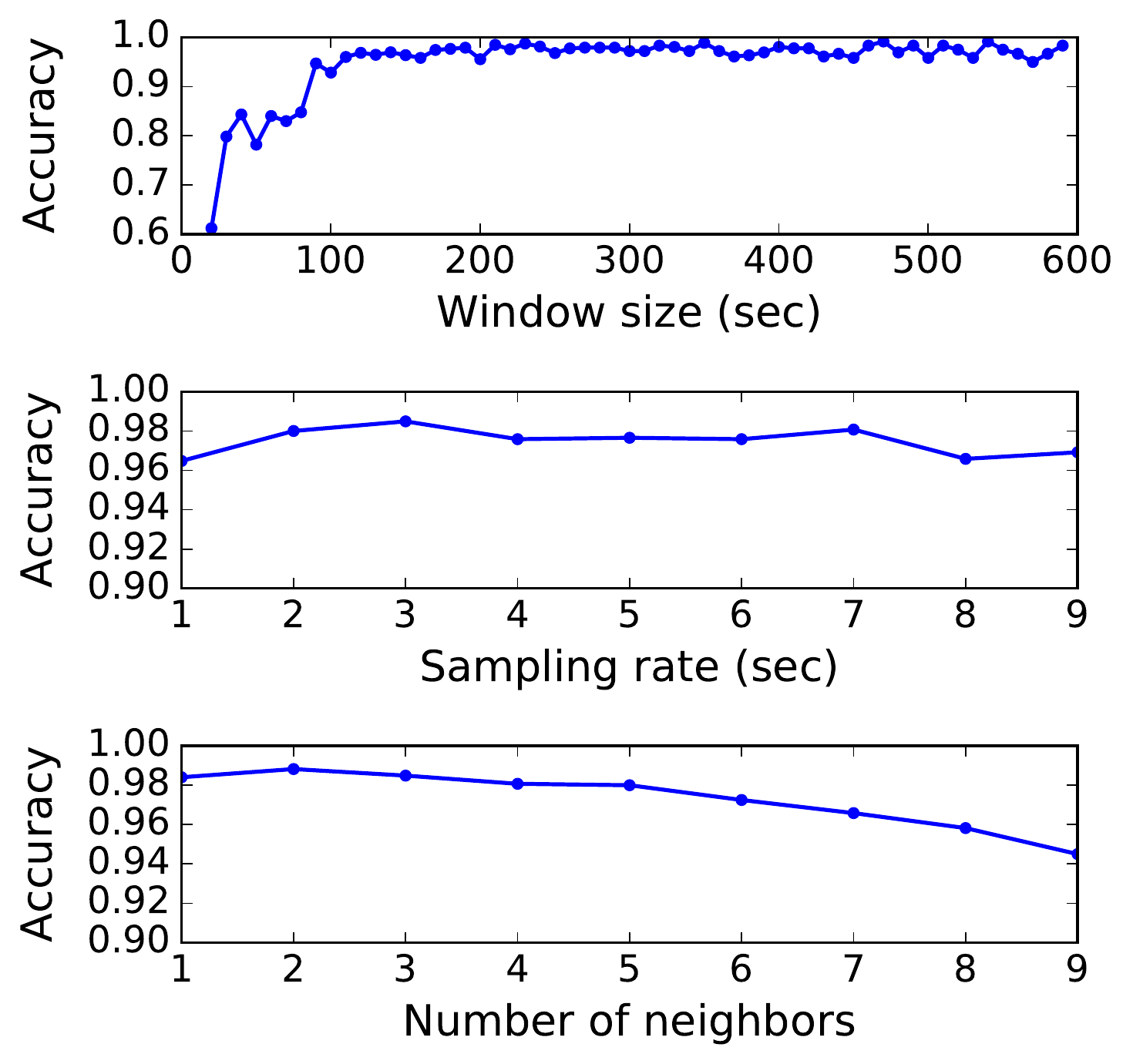}
\caption{Simple supervised machine learning can perform high-accuracy device identification from smart home traffic rates. These plots show the accuracy of a $k$-nearest-neighbors classifier identifying which of six commercially-available smart home devices generated the traffic within a time window for varying window size, sampling rate, and number of neighbors $k$.  Accuracies determined by 10-fold stratified cross-validation.}
\label{fig:classification}
\end{figure}

A $k$-nearest-neighbors classifier trained on these feature vectors resulted in greater than $95\%$ accuracy classifications for a range  of $w$, $s$, and $k$ (Figure~\ref{fig:classification}). 
Accuracies were determined by 10-fold stratified cross-validation on the training set. 
The success of such simple machine learning methods indicates that an adversary, especially one with more data and more sophisticated models, could likely use machine learning to identify devices from consumer data. 

This approach does have a drawback: it requires the observer to obtain training data for every IoT device that the observer wishes to potentially identify. 
Given the number of new IoT devices entering the market, it is likely that only well-resourced observers could maintain a sophisticated enough training set for high accuracy classification  from all smart homes. 
Further, it is possible that it will be harder to distinguish devices based on our chosen features as more devices enter the market.
Future research focused on using machine learning in this domain will be necessary to support or refute our initial findings that devices are identifiable from traffic rate metadata alone. 
Until then, concealing DNS significantly raises the barrier to device identification for last-mile observers.

The smart home domain further motivates the need for encrypted DNS, for which there are already proposals \cite{dnsrfc, dnscrypt}.

\subsection{Tunneling traffic}
Another strategy for preventing device identification and consumer behavior inference is to tunnel all smart home traffic through a virtual private network (VPN).
This would be done to impede a last-mile observer's ability to split traffic into individual streams and associate those streams with specific devices.

A VPN wraps all traffic from an endpoint (like a home gateway router) in an additional transport layer, aggregating it into a single stream.
This stream has the source IP address of the home gateway router and the destination IP address of the server implementing the VPN exit point.
In effect, a adversary would see all traffic as originating and terminating from a single pair of endpoints, rather than from individual smart home devices and their cloud servers.

We evaluated the effectiveness of a VPN in preventing traffic-splitting. 
We removed all information from our recorded IoT traffic data other than network traffic rates aggregated over all devices.
We have yet to find a general method of associating traffic back to individual devices without knowing the number of devices or using categorical header information. 
When many IoT devices send traffic simultaneously, noise variations in traffic rates of high-rate devices can mask the entire traffic patterns of other devices. 
This makes it difficult to determine which variations in the overall traffic rate observed from outside a VPN correspond to individual devices. 

However, it is possible that in specific, limited circumstances a dedicated observer could infer individual devices' traffic even when tunneled over a VPN.
If there are multiple devices that send traffic at different times, supervised machine learning could be used to identify devices because most time windows would contain traffic from only a single device. Even if device traffic overlaps, an observer could possibly infer the presence of the device that sends the most traffic if it significantly overshadows the traffic from other devices.

Outside of these specific cases, the wider deployment of VPNs into smart homes would be an effective solution to the traffic rate privacy threat from last-mile observers.
However, there are several drawbacks to relying on VPNs as the sole solution.
Properly setting up a home VPN is difficult for the typical consumer  \cite{upturn}. 
This could make the cost of widespread VPN deployment prohibitively high.
A well-designed smart home hub could make deployment easier.
If the hub acted as a WiFi access point for multiple smart home devices, it could come pre-configured as a VPN and be relatively easy for consumers to use.

Using a VPN does not prevent the VPN exit node's last-mile provider from performing device identification and behavior inference.
This problem could be addressed by having the VPN provider act as an endpoint for traffic from multiple smart homes.
In that case, it would be difficult for the VPN's last-mile provider to associate devices with a particular smart home. 
We believe that such ``mixes" for smart homes could work in a similar fashion as anonymity mixes in other contexts \cite{chaum1981untraceable, dingledine2004tor}.

Using a VPN to prevent device identification and behavior inference has no effect on the WiFi eavesdropper. 
WiFi transmissions from individual devices can still be intercepted independently of a hub or gateway router using a VPN.

\subsection{Shaping and injecting traffic}
The previous strategies predominantly address the traffic rate privacy threat in the context of last-mile observers.
WiFi eavesdroppers present a greater difficulty since they are able to directly measure radio traffic coming to and from each smart home device.

In the case of the WiFi eavesdropper, it is therefore necessary to shape or inject traffic to reduce the observer's confidence in device identification or behavior inference.
Shaping or injecting traffic would also help protect against a last-mile observer when a VPN is not entirely effective at preventing traffic disaggregation.
Solutions involving traffic shaping or injection could be implemented by device manufacturers or by a third-party on a hub or gateway router.

\subsubsection{Device manufacturer implementation}
Developers could implement devices such that network traffic doesn't directly correspond to user behavior in real-time.
For some applications this make sense. 
A sleep monitor, for instance, does not need to update a cloud service immediately upon detecting that a user has gone to sleep---it could delay the notification for up to several hours without sacrificing functionality.
Introducing random delays would reduce an observer's confidence in inferred sleep patterns.
However, some devices rely on real-time updates. 
A smart personal assistant cannot wait before answering a user's questions.

A device and its cloud service could also inject encrypted traffic to maintain a constant traffic rate regardless of consumer interactions.
This would make idle device behavior indistinguishable from interesting consumer behaviors.
Since data caps are a concern for consumers, maintaining a constant, high traffic rate is not a viable solution \cite{odlyzko2012know}. 

Alternatively, a device could send decoy network traffic when no consumer interaction has occurred.
If done correctly, this would lower an observer's confidence in any behavior inference.
A device manufacturer with access to the device's software can trigger fake interactions indistinguishable from real events.

\emph{When} to inject traffic for fake consumer events depends on the typical behavior of the consumer.
It is important that fake events happen at logical times (i.e. a smart personal assistant shouldn't answer a question after the user has gone to sleep) and that their distribution reflect plausible consumer behaviors.
Doing so is a difficult problem, which we address below.

\subsubsection{Third party implementations}
Traffic spoofing solutions could also be implemented on a third-party hub or gateway router, which could protect multiple devices simultaneously.

The hub or router would first need to identify what traffic patterns correspond to consumer behaviors.
Unlike device developers, a third-party hub or router would have to infer what traffic patterns map to consumer interactions for each device---effectively having to perform the very behavior inference they are trying to prevent. 
For example, a hub or router would need to determine that an interaction with a smart outlet generates a short, sharp spike in network traffic while an interaction with a smart security camera creates an longer period of high traffic volume.
There is also indeterminancy in device traffic patterns from real consumer interactions, so a hub or router would need to generate a model to introduce realistic noise into injected traffic.  

For some devices, even deciding what constitutes a discrete consumer interaction is an open question.
For example, a fitness tracker records a consumer's individual footsteps.
A set of steps could constitute a ``walk to the car."
Should a third-party hub or router attempt to mimic individual steps, or a logical walk?
With such device-dependent consumer interactions, identifying which traffic patterns to mimic in a robust, automated way is an open research problem.

A second difficulty of spoofing traffic for third-party hubs and routers is deciding \emph{when} to initiate decoy interactions.
This is also a difficulty for device manufacturers, though they may have better models for how consumers use their proprietary devices.
Conceptually, decoy interactions should function like ``parallel realities.''
The goal of injecting decoy actions is to make it difficult for an observer to choose which reality is the correct one.
The distribution of decoy events every day must therefore resemble the distribution of real events.
Consider a consumer who owns a smart home door lock that sends traffic indicating when the consumer has returned home from work.
Well-formed decoy traffic would not create a parallel reality in which the consumer interacted with a personal assistant device located in their home \emph{before} they returned home from work.
Creating logically-sound parallel realities is difficult, especially as the number of devices increases.

Finally, after deciding what to mimic and when, a third-party hub or router has to decide \emph{how} it will implement decoy injection.
Replaying recorded TCP connections is impossible because they are encrypted.
If a third-party attempts to forge a new connection with a cloud server, the resulting traffic profile is unlikely to resemble a real device's traffic pattern \cite{houmansadr2013parrot}.

A VPN helps overcome this obstacle, since decoy traffic can be sent to the VPN endpoint rather than a device's cloud service.
As long as the VPN endpoint cooperates, the router or hub can exchange spurious communication with the VPN endpoint acting as the cloud server.
To a last-mile observer, the traffic between the home and the VPN endpoint would appear to contain meaningful consumer actions when no meaningful communication has taken place.
This would limit possible behavior inference by a last-mile observer in the special cases when a VPN by itself is insufficient.

However, sending decoy traffic over a VPN still does not protect against a WiFi eavesdropper because no additional 802.11 frames are generated---all traffic is going out on the wire.
A third-party hub or gateway router must also forge decoy WiFi frames with MAC addresses spoofed to match the addresses of the smart home devices.

There are sophisticated methods a WiFi eavesdropper could employ to discover which frames have spoofed MAC addresses.
The origin of WiFi radio signals can be triangulated to their source using signal strengths recorded from a multi-antenna array \cite{xiong2013arraytrack}.
Therefore, even if a third-party hub or router perfectly mimics a device over WiFi, it is possible that a sophisticated enough eavesdropper could deduce which radio traffic is fake.

Though difficult to implement, traffic spoofing and injection would offer probabilistic guarantees of safety against traffic metadata privacy threats.
We are currently researching how to shape or inject smart home traffic to provably protect consumer privacy while minimizing the amount of additional traffic needed.

\section{Discussion}
We have outlined several technical strategies to limit an observer's ability to identify smart home devices and infer consumer behavior. While these strategies will require further research before they are ready for consumer use, we make the following specific recommendations:
\begin{itemize}
\item Developers should consider how their devices default to a ``minimum reliable product'' in the face of limited Internet connectivity.
\item The Internet ecosystem should work towards an encrypted DNS protocol.
\item Smart home hubs and gateway routers should provide consumers with the ability to easily configure their devices to use VPNs.
\item Researchers should continue to explore traffic shaping techniques that would limit an adversary's ability to infer sensitive consumer information from traffic rate metadata.
\end{itemize}

Policy measures could also play a role in enacting these and other suggestions for protecting consumers against the improper use of traffic rate metadata. 
For example, policymakers should push device manufacturers to ensure that their devices default to a ``minimum reliable product" as described above. This would allow consumers to enjoy some of the benefits of  smart devices while selectively limiting outgoing traffic from those devices. 

Policymakers could also regulate how actors like Internet service providers collect and use traffic rate information.
The Federal Communications Commission enacted new rules in 2016 that impacted how Internet Service Providers could collect and use data \cite{fcc-nprm}, however, these rules seem likely to change in the coming months.  Future regulation that gives consumers transparency and control over the collection and use of smart home traffic data would help address the issues we have presented and alleviate privacy concerns.

\section{Conclusion}
The side-channel privacy threat of traffic rate metadata will continue to grow along with the market for IoT smart home devices.
In this paper, we proposed and evaluated four strategies to protect against this threat, including blocking traffic, concealing DNS, tunneling traffic, and shaping and injecting traffic. 
While there is unlikely to be a single ``silver bullet'' solution, each of our proposed strategies would be effective under specific circumstances and provides a framework for  further development.  
We hope that this discussion assists  researchers, policymakers, and device manufacturers attempting to create privacy solutions and contributes to the growing interest in the privacy implications of smart home IoT devices.
\label{sec:conclusion}

\section*{Acknowledgements}
Thanks to Arvind Narayanan and Margaret Martonosi. This research was funded by the Department of Defense through the National Defense Science \& Engineering Graduate Fellowship (NDSEG) Program, a Google Faculty Research Award, and the National Science Foundation.
\bibliographystyle{IEEEtran}
\bibliography{ClosingTheBlinds}

\end{document}